\documentclass[10pt,final,twocolumn,journal]{IEEEtran}
\usepackage{url}
\usepackage[noadjust]{cite}
\usepackage{epstopdf}
\usepackage{cite}
\usepackage{graphicx}
\usepackage{caption}
\usepackage{subcaption}
\usepackage{authblk}
\usepackage{xcolor}
\usepackage{amsmath}
\usepackage{array}
\newcolumntype{P}[1]{>{\centering\arraybackslash}p{#1}}
\usepackage{enumitem}
\usepackage{booktabs}
\usepackage{adjustbox}
\usepackage[font=small,labelfont=bf]{caption}

\begin{document}
\title{Channel thickness optimization for ultra thin and 2D chemically doped TFETs}
\author{Chin-Yi Chen, Tarek A. Ameen, Hesameddin Ilatikhameneh, Rajib Rahman, Gerhard Klimeck, Joerg Appenzeller\vspace{-6.5ex}
\thanks{This work was supported in part by the Center for Low Energy Systems Technology (LEAST), one of six centers of STARnet, a Semiconductor Research Corporation program sponsored by MARCO and DARPA.}
\thanks{The authors are with the Department of Electrical and Computer Engineering, Purdue University, West Lafayette, IN, 47907 USA e-mail: chen1648@purdue.edu}
}
\maketitle

\setlength{\textfloatsep}{12pt} 
\setlength{\belowdisplayskip}{1.6pt} 
\setlength{\belowdisplayshortskip}{1.6pt}
\setlength{\abovedisplayskip}{1.6pt} 
\setlength{\abovedisplayshortskip}{1.6pt}
\setlength{\belowcaptionskip}{-12pt}
\vspace{-1.0\baselineskip}
\begin{abstract}
2D material based tunnel FETs are among the most promising candidates for low power electronics applications, since they offer ultimate gate control and high current drives that are achievable through small tunneling distances ($\Lambda$) during the device operation. The ideal device is characterized by a minimized $\Lambda$. However, devices with the thinnest possible body do not necessarily provide the best performance. For example, reducing the channel thickness ($T_{ch}$) increases the depletion width in the source which can be a significant part of the total $\Lambda$. Hence, it is important to determine the optimum $T_{ch}$ for each channel material individually. In this work, we study the optimum $T_{ch}$ for three channel materials: WSe$_{2}$, Black Phosphorus (BP), and InAs using full-band self-consistent quantum transport simulations.  To identify the ideal $T_{ch}$ for each material at a specific doping density, a new analytic model is proposed and benchmarked against the numerical simulations.
\end{abstract}

\section{Introduction}

\begin{figure}[!b]
\includegraphics[width=3.0in]{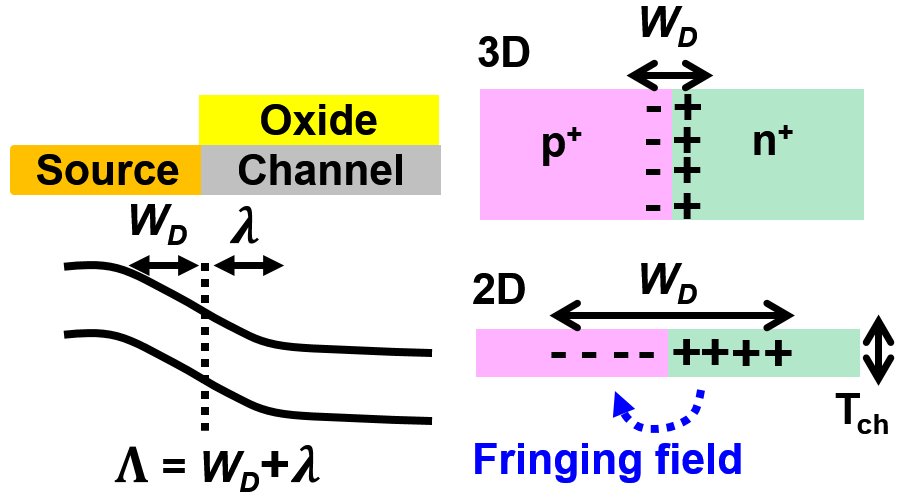}
\caption{TFET's tunneling distance ($\Lambda$) contains a depletion width ($W_{D}$) in the source region and the scaling length ($\lambda$) in the channel. $W_{D}$ of the 2D PN junction is larger than $W_{D}$ of the 3D PN junction due to the presence of fringing fields.}
\label{schematic_CD_TFET_Lambda}
\end{figure}

The supply voltage of metal oxide semiconductor field-effect transistors (MOSFETs) has almost stopped scaling since the beginning of the millennium, since the sub-threshold swings in conventional transistors cannot be imporved beyond 60 mV/dec at room-temperature for fundamental reasons. Even today's state-of-the-art 14nm tri-gate MOSFET does not operate below 0.6V \cite{Bohr2011}. For decades, the semiconductor industry has been researching transistors with steeper subthreshold characteristics $I_{D} - V_{G}$ than obtainable in conventional FETs with the goal to enbale low supply voltages to reduce power consumption \cite{itrs2015,Gonzalez1997}. Unlike MOSFETs, tunneling field effect transistors (TFETs) are not bound by the 60mV/dec limit and can operate at significantly lower voltages \cite{Appenzeller_2004,Appenzeller_2005,choi_2007,Ionescu_2011}. Such merit makes TFETs an appealing alternative option to MOSFETs for low power applications \cite{Seabaugh2010,Avci2015}. However, TFETs utilize band to band tunneling (BTBT) to switch the device ON and OFF, and the BTBT process limits the ON current ($I_{ON}$) accordingly. 

$I_{ON}$ is proportional to the BTBT transmission probability ($T_{BTBT}$) which can be expressed in terms of both electrostatics and material properties \cite{Hesam_FN_tunneling} as 
\begin{align}\label{eq:T_BTBT}  
     I_{ON} \propto T_{BTBT} \propto e^{-\Lambda\sqrt{m^{*}E_{g}}}, 
\end{align}
where in the simplest picture $m^*$ is the reduced effective mass along the transport direction ($\frac{1}{m_{}^{*}}=\frac{1}{m_{e}^{*}}+\frac{1}{m_{h}^{*}}$). $E_{g}$ is the band gap of the channel material and $\Lambda$ is the tunneling distance at the junction. Reducing $\Lambda \sqrt{m^{*}E_{g}}$ increases $T_{BTBT}$ and $I_{ON}$ exponentially. For a more detailed discussion on equation (\ref{eq:T_BTBT}) and its accuracy, please refer to Appendix I.

For a chemically doped TFET, the total tunneling distance ($\Lambda$) has two contributions \cite{Salazar2015,Hesam_FN_tunneling}: the depletion width ($W_{D}$) in the doped source region and the scaling length ($\lambda$) in the channel as shown in Fig. \ref{schematic_CD_TFET_Lambda}. 

 \begin{figure}[!b]
\centering
\includegraphics[width=2.2in]{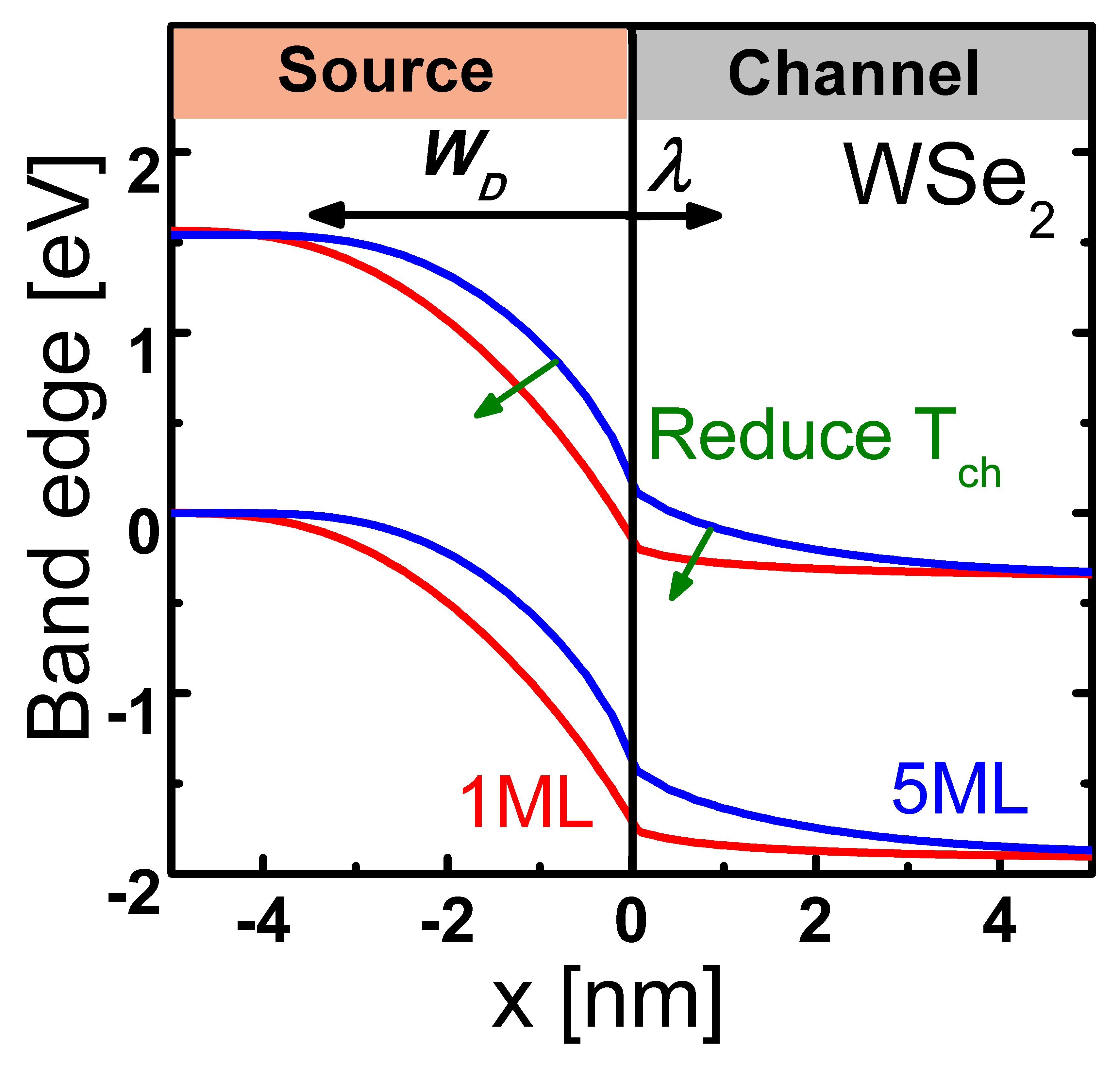}
\caption{Band diagrams of a mono-layer and a 5ML WSe$_{2}$ TFET extracted from atomistic quantum transport simulations.}
\label{WSe2_ECEV}
\end{figure}

Reducing the channel thickness ($T_{ch}$) is beneficial in terms of electrostatics and may translate into a smaller $\Lambda$ \cite{Avci2015} if for example $W_{D}$ does not depend on $T_{ch}$. In reality, however, $W_{D}$ is thickness dependent in a low dimensional system (i.e. 2D material). This has been demonstrated experimentally \cite{2D-PN-experiment}, numerically, and analytically \cite{Hesam_pn_junction,2D-PN-experiment,Nipane2017,Yu2016} where $W_{D}$ is larger in a 2D compared to a 3D PN junction. In a 2D PN junction, $W_{D}$ is inversely proportional to the thickness \cite{Hesam_pn_junction}.
\begin{align}    
      W_{D} & = \dfrac{\pi \varepsilon \Delta V}{qNT_{ch}}  \label{eq:WD}        
\end{align}
where $N$ is the doping density and $\Delta V$ is the built-in potential. $\varepsilon$ is an averaged dielectric constant of the channel material and the dielectric surrounding the source.
$\lambda$ for a chemically doped double gated 2D TFET \cite{Hesam_DETFET} can be approximated as
\begin{align}
          \lambda \quad & = \dfrac{\varepsilon}{\varepsilon_{ox}}\left[\gamma_1 T_{ch} + \gamma_2 T_{ox} \right] \label{eq:screening_length}              
\end{align}
where $\varepsilon_{ox}$ and $T_{ox}$ are the dielectric constant and the thickness of the gate oxide in Fig. \ref{schematic_CD_TFET_Lambda}, respectively. $\gamma_1$ and $\gamma_2$ are fitting parameters, since the expression without $\gamma_1$ and $\gamma_2$ was derived for an electrostatically doped 2D TFET \cite{Hesam_ED_TFET,Hesam_DETFET}.

\begin{figure}[!t]
\centering
\includegraphics[width=2.0in]{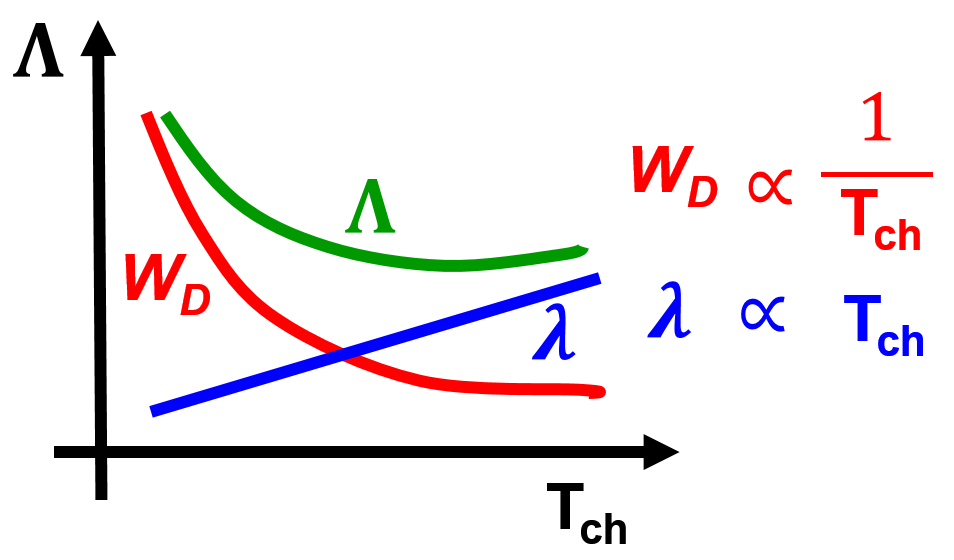}
\caption{Scaling length ($\lambda$) is proportional to $T_{ch}$ while depletion width ($W_{D}$) is inversely proportional to $T_{ch}$. As a result, the thinnest possible $T_{ch}$ does not guarantee the smallest $\Lambda$.}
\label{schematic_Lambda}
\end{figure}

Fig. \ref{WSe2_ECEV} shows that $\lambda$ and $W_{D}$ respond to $T_{ch}$ in the opposite fashion as illustrated in the band diagrams extracted from atomistic simulations of a mono-layer (1ML) and a 5ML WSe$_{2}$ TFETs. When $T_{ch}$ is reduced from 5ML to 1ML, $\lambda$ is reduced due to the tighter gate control according to eq. (\ref{eq:screening_length}) while $W_{D}$ increases according to eq. (\ref{eq:WD}). As a result, the thinnest possible $T_{ch}$ may not minimize $\Lambda$ as shown in Fig. \ref{schematic_Lambda}.

Moreover, the channel thickness ($T_{ch}$) that minimizes the total tunneling distance ($\Lambda$) is not necessarily the best $T_{ch}$ overall since material parameters might change also with $T_{ch}$. To obtain the highest $I_{ON}$, the optimum  $T_{ch}$ should minimize the entire expression $\Lambda \sqrt{m^{*}E_{g}}$. In this work, a compact model to optimize $T_{ch}$ for the ON state ($T_{ch.opt}$) is introduced, and the model is benchmarked with state-of-the-art atomistic quantum transport simulations.

Materials considered in this article can be classified depending on how their band gap changes with $T_{ch}$. We define class I materials as those that do not show a dependence of $E_{g}$ on $T_{ch}$, which, as will be discussed below, results in thinner $T_{ch.opt}$ for optimum ON-state performance. On the other hand, class II materials, according to our definition, exhibit a decrease of $E_{g}$ with increasing $T_{ch}$. As a result, an optimum design is achieved with a relatively thicker $T_{ch.opt}$. Details are described later.

\begin{figure}[!t]
\centering
\includegraphics[width=2.0in]{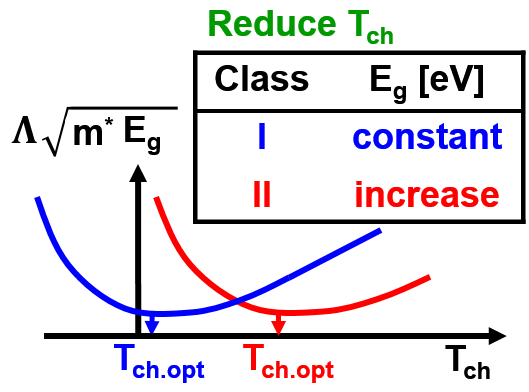}
\caption{ An optimum channel thickness ($T_{ch.opt}$) exists that minimizes $\Lambda \sqrt{m^{*}E_{g}}$. Depending on whether or not $E_{g}$  changes with the channel thickness ($T_{ch}$), the optimum body thickness ($T_{ch.opt}$) occurs at smaller or larger $T_{ch}$ respectively. }
\label{schematic_gamaLEg}
\end{figure}

In sections II and III, the impact of the body thickness on material properties and tunneling distance are discussed. Section IV shows the optimized $T_{ch}$ for the ON-state obtained from an analytic analysis and atomistic quantum transport simulations. Section V shows the upper limit of $T_{ch}$ in a TFET. Last, section VI summarizes the $T_{ch}$ design rules for a TFET.

\section{The impact of $T_{ch}$ on $E_{g}$ and $m^*$}

Three channel materials are considered in this work: WSe$_{2}$, BP, and InAs. BP and InAs are chosen, because their small direct $E_{g}$ occurs promising for TFET applications \cite{Hesam_moors_law,Beneventi2014,Ameen2016,Ford2011}. Note that in the case of InAs, body thicknesses beyond what has been experimentaqlly achieved were considered, and that in general transport in channels with $T_{ch}$ below 5 nm is strongly impacted by surface scattering \cite{Tomioka2012,Cutaia2016}, while our model assumes that ballistic transport conditions prevail. WSe$_{2}$ is chosen as a case study, since its direct band gap ($E_{g}$) barely changes with $T_{ch}$. Moreover, WSe$_{2}$ based TFETs are expected to show the best performance
among semiconducting transition metal dichalcogenides (TMDs)  \cite{Hesam_TFETs_2D_TMD,Hesam_design_rule_2DTFET,Ma2015}.

\begin{figure}[!h]
\centering
\includegraphics[width=1.6in]{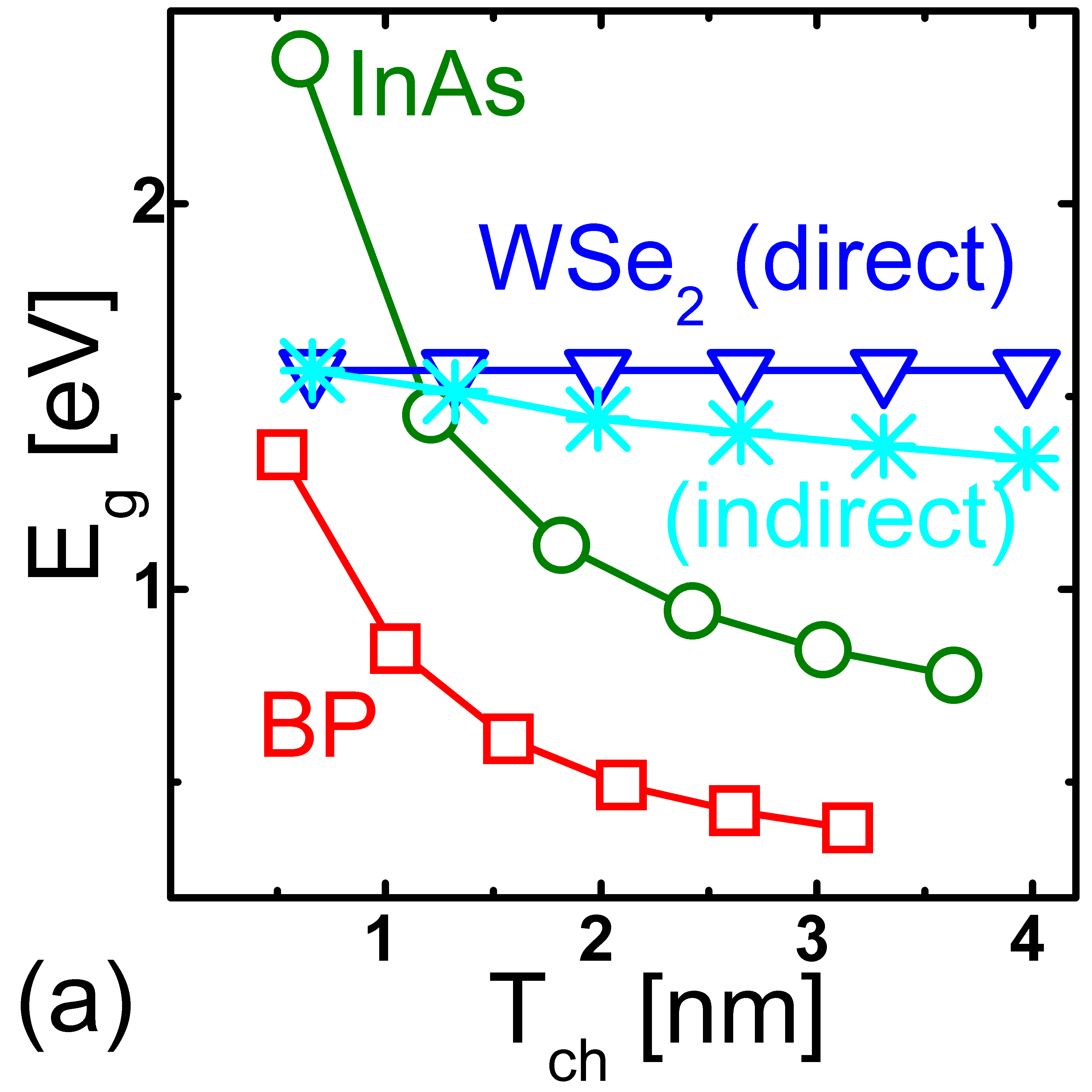}
\includegraphics[width=1.6in]{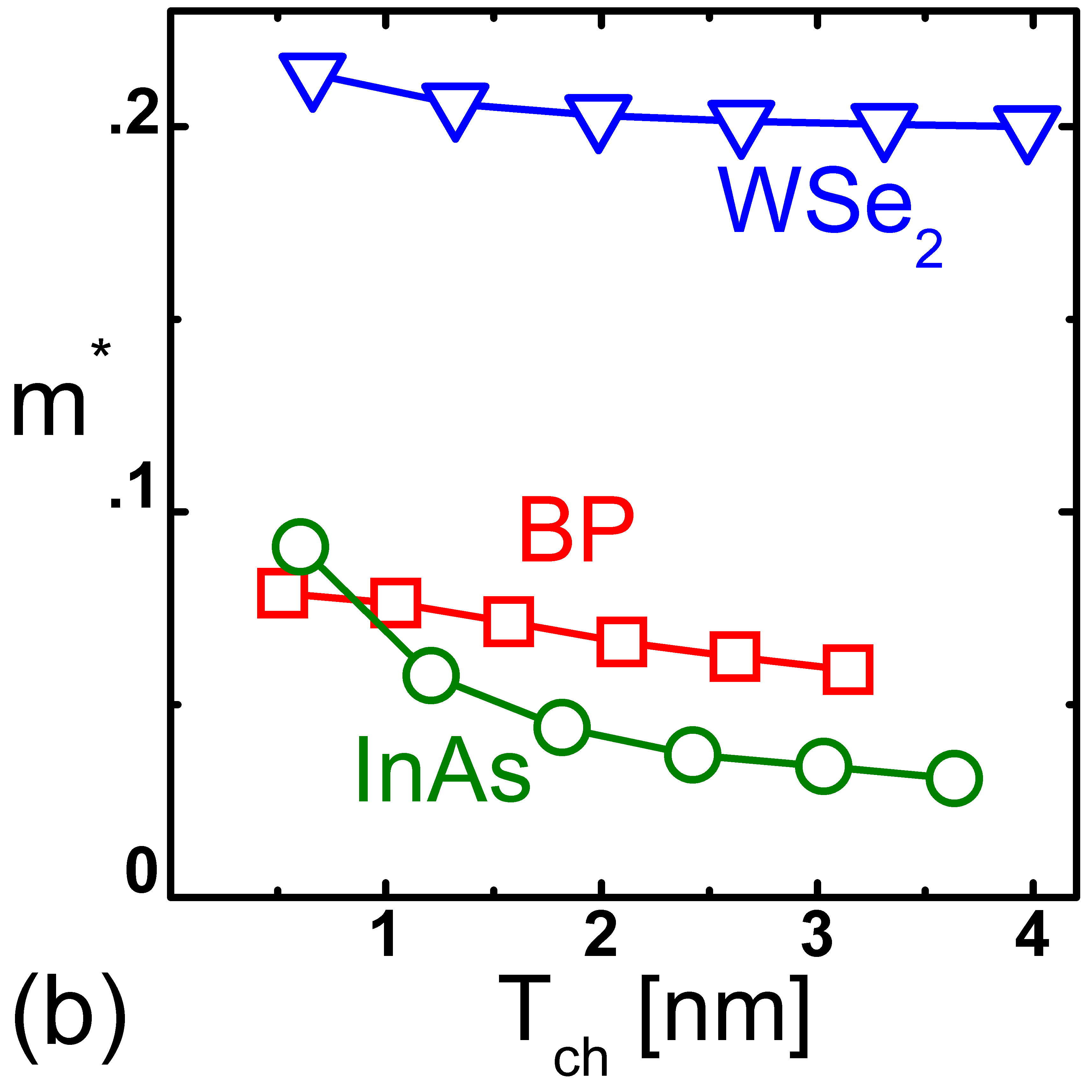}
\caption{(a) and (b) show three studied channel materials' $E_{g}$ and tunneling mass $m^*$ as a function of $T_{ch}$ extracted from atomistic tight binding simulations, which are calibrated against density function theory (DFT).}
\label{material_property}
\end{figure}	

Fig. \ref{material_property} shows $E_{g}$ and tunneling mass $m^{*}$ of WSe$_{2}$, BP, and InAs extracted from atomistic tight binding simulations. The Slater-Koster tight binding parameters \cite{Koster1954} of WSe$_{2}$ are extracted from the band structure calculated by density function theory (DFT) with generalized gradient approximation (GGA) \cite{Hesam_TFETs_2D_TMD,Gong2013}. Note that $E_{g}$ is not exactly the same as in some transport experiments \cite{Prakash_2017}, but is comparable.

$E_{g}$ and $m^*$ typically increase with stronger confinement achieved by reducing $T_{ch}$. The dependence of $E_{g}$ and $m^{*}$ on $T_{ch}$ can be expressed as $E_{g}=E_{g.bulk}+\frac{\alpha}{T_{ch}}$ and $m^{*}=m^*_{bulk}+\frac{\alpha^{'}}{T_{ch}}$. $\alpha$ and $\alpha^{'}$ are determined from Fig. \ref{material_property} by fitting. The parameter $\alpha$ for the case of WSe$_{2}$ is significantly smaller than in the case of BP and InAs due to weak inter-layer coupling \cite{Novoselov2016,Wang2017}.


\begin{table} [!htbp]
\begin{center}
\rule{0pt}{8pt}
\begin{tabular}{ |c|cccc| } 
 \hline
 & & & & \\ [-1em]
 channel    & $E_{g.bulk}$  & $\alpha$  & $m^{*}_{bulk}$ &  $\alpha^{'}$ \\ 
            & [eV]          & [eV/nm]   & [$m_{0}$]      & [$m_{0}/nm$] \\ 
 \hline
  & & & & \\ [-1em]
 WSe$_{2}$  & 1.535     &  0.02   & 0.2    &  0.01    \\ 
  & & & & \\ [-1em]
 BP         & 0.28    &  0.52   & 0.055    &  0.02    \\ 
  & & & & \\ [-1em]
 InAs       & 0.35    &  1.5   & 0.025    &  0.035   \\ 
 \hline
\end{tabular}
\end{center}
\caption{Parameters $E_{g.bulk}$, $m^{*}_{bulk}$, $\alpha$, and $\alpha^{'}$ for WSe$_{2}$, BP, and InAs. }
\label{table:Eg_m}
\end{table}

 \begin{figure}[!h]
\centering
\includegraphics[height=1.14in]{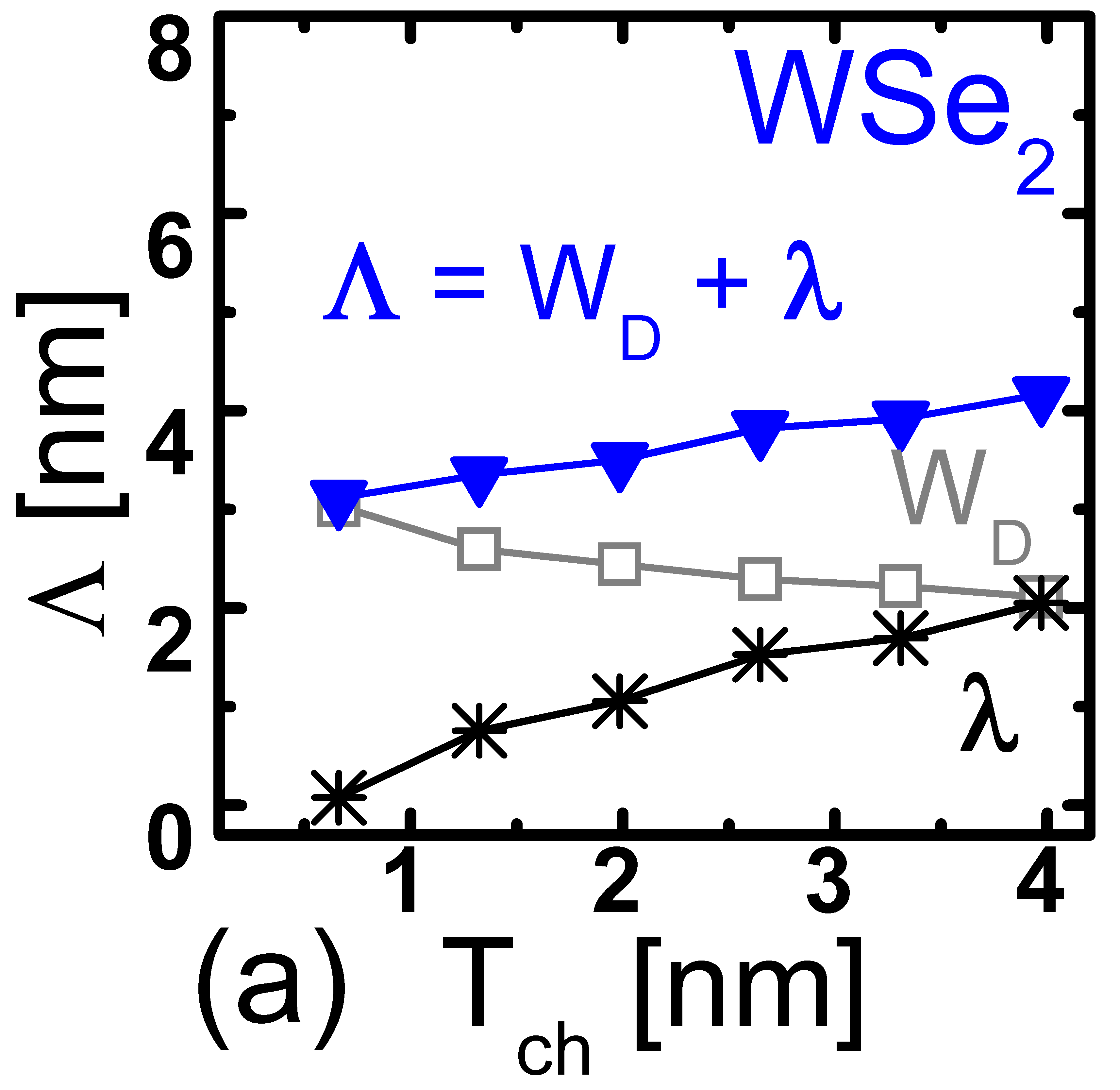}
\includegraphics[height=1.14in]{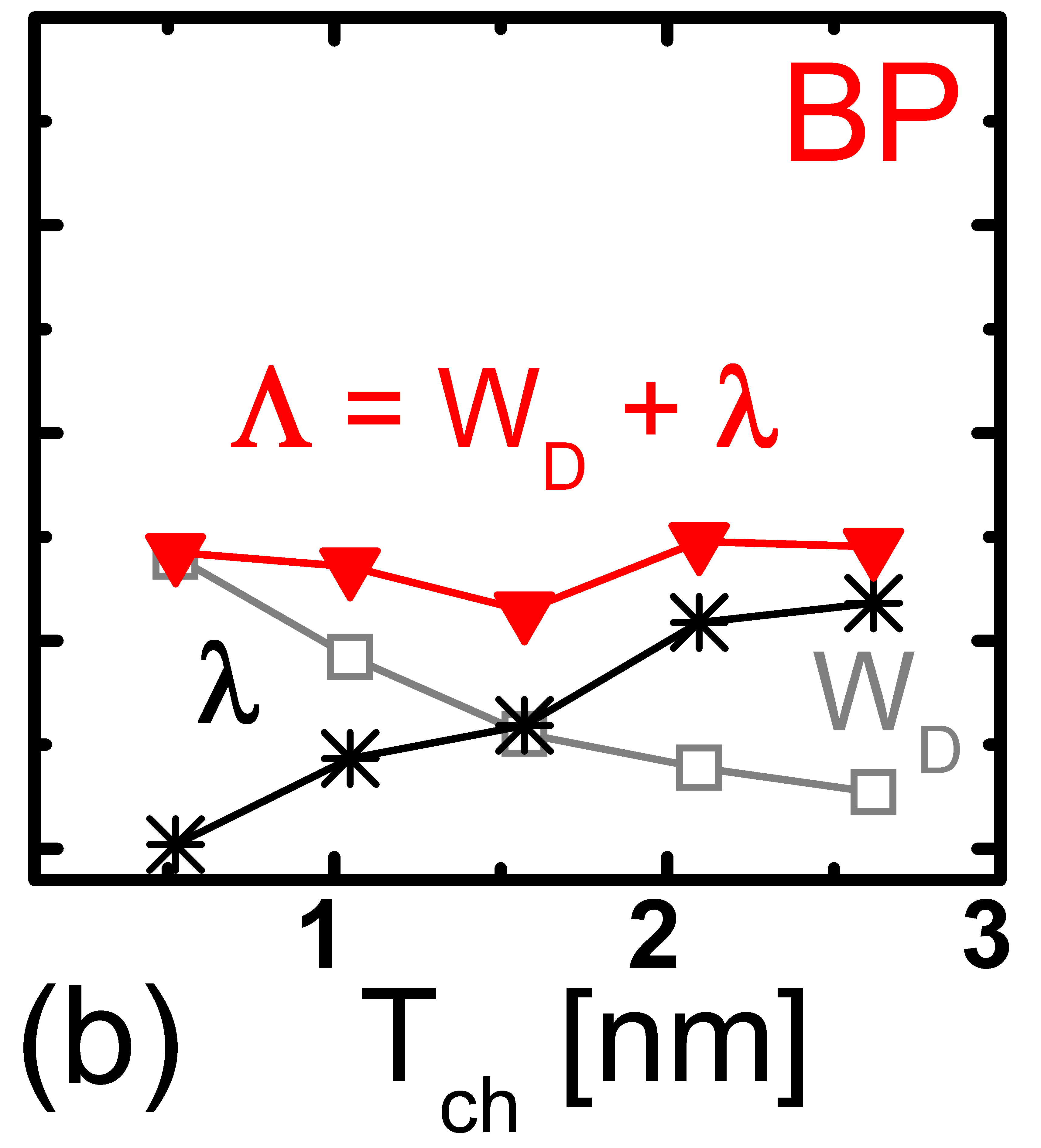}
\includegraphics[height=1.14in]{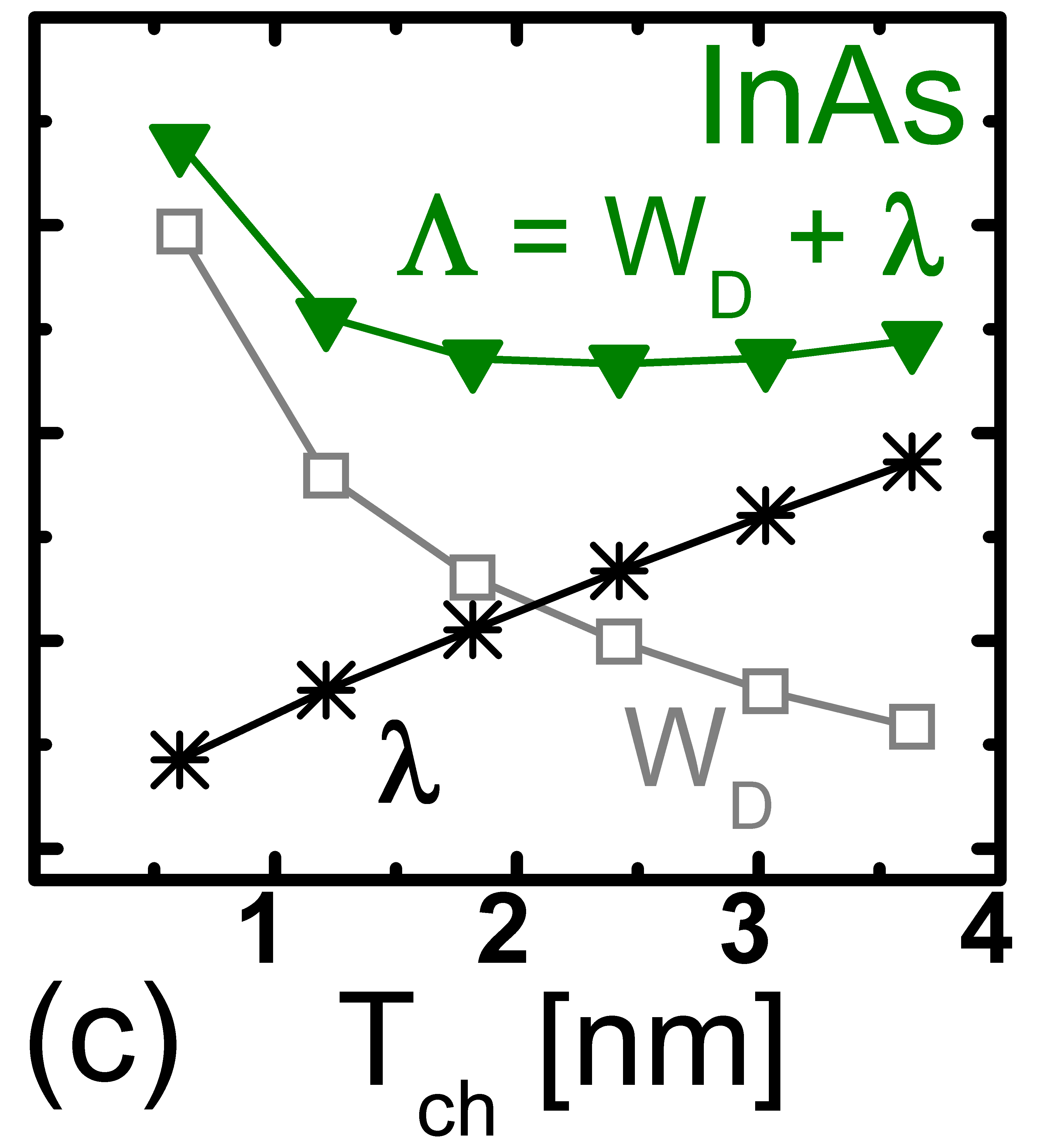}
\caption{(a), (b), and (c) Dependence of $\Lambda$, $\lambda$, and $W_{D}$ on $T_{ch}$ for WSe$_{2}$, BP, and InAs TFETs, respectively.}
\label{tunnel_distance}
\end{figure}

\section{The impact of $T_{ch}$ on the tunneling distance}

WSe$_{2}$, BP, and InAs 2D TFETs' $\lambda$, $W_{D}$, and $\Lambda$ are shown in Fig. \ref{tunnel_distance} (a), (b), and (c). The source doping density (N) is $10^{20}\ cm^{-3}$. All simulated materials show that $\lambda$ is proportional to $T_{ch}$ while $W_{D}$ is inversely proportional to $T_{ch}$. The data are extracted from atomistic quantum transport simulations. The details of the simulation method and the simulated structure is described in Appendix II. $W_{D}$ in the ON-state is proportional to the potential drop ($\Delta V$) across the depletion layer in the source, which is of the order of $E_{g}$. Since $E_{g}$ of of BP and InAs has a stronger dependence on $T_{ch}$ compared to WSe$_2$, their $W_{D}$ also shows a stronger dependence on $T_{ch}$.

\section{The Optimum $T_{ch}$ for ON-state ($T_{ch.opt}$) }
\begin{figure}[!b]
\centering

\includegraphics[width=2.2in]{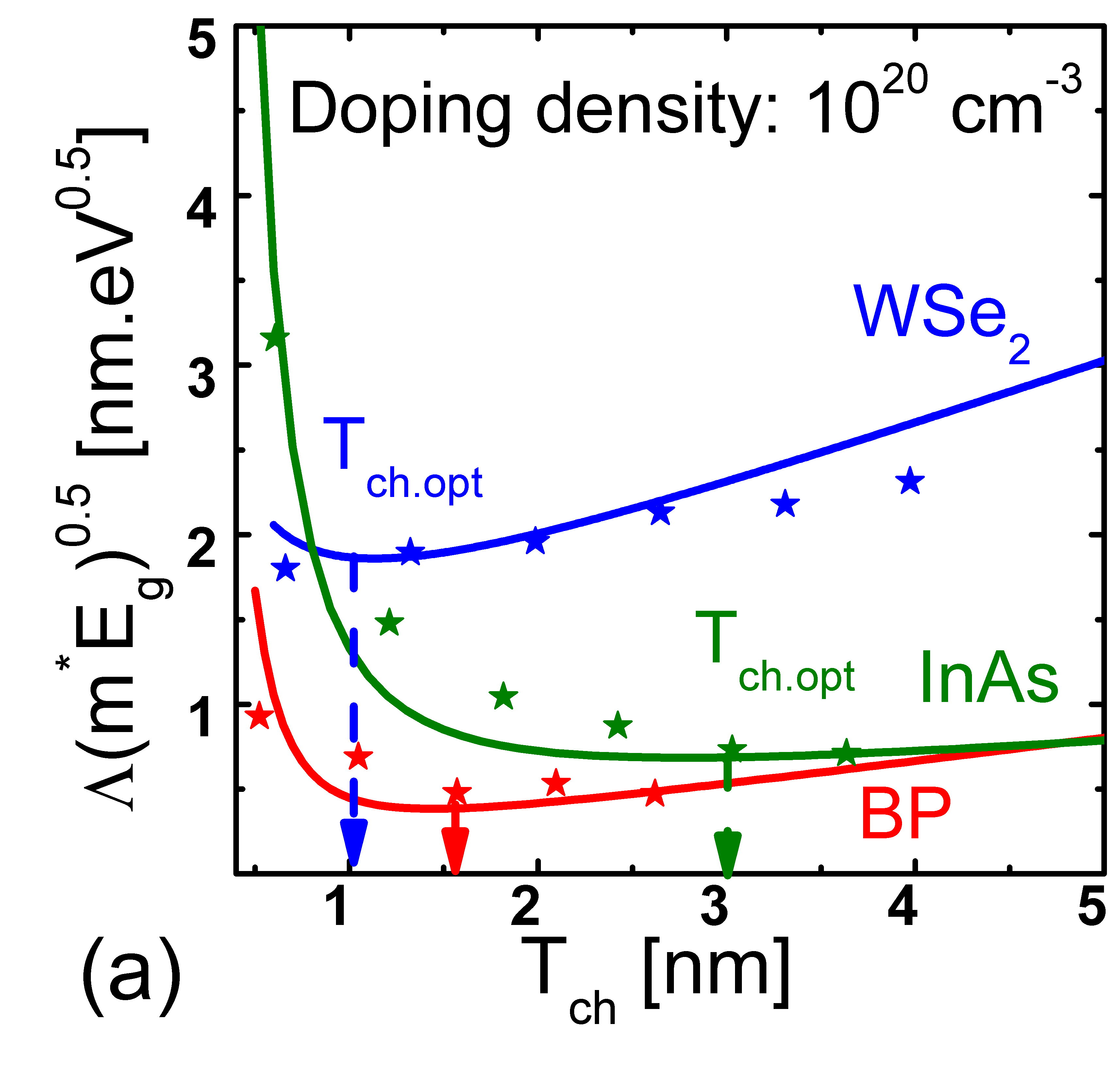}
\includegraphics[width=2.2in]{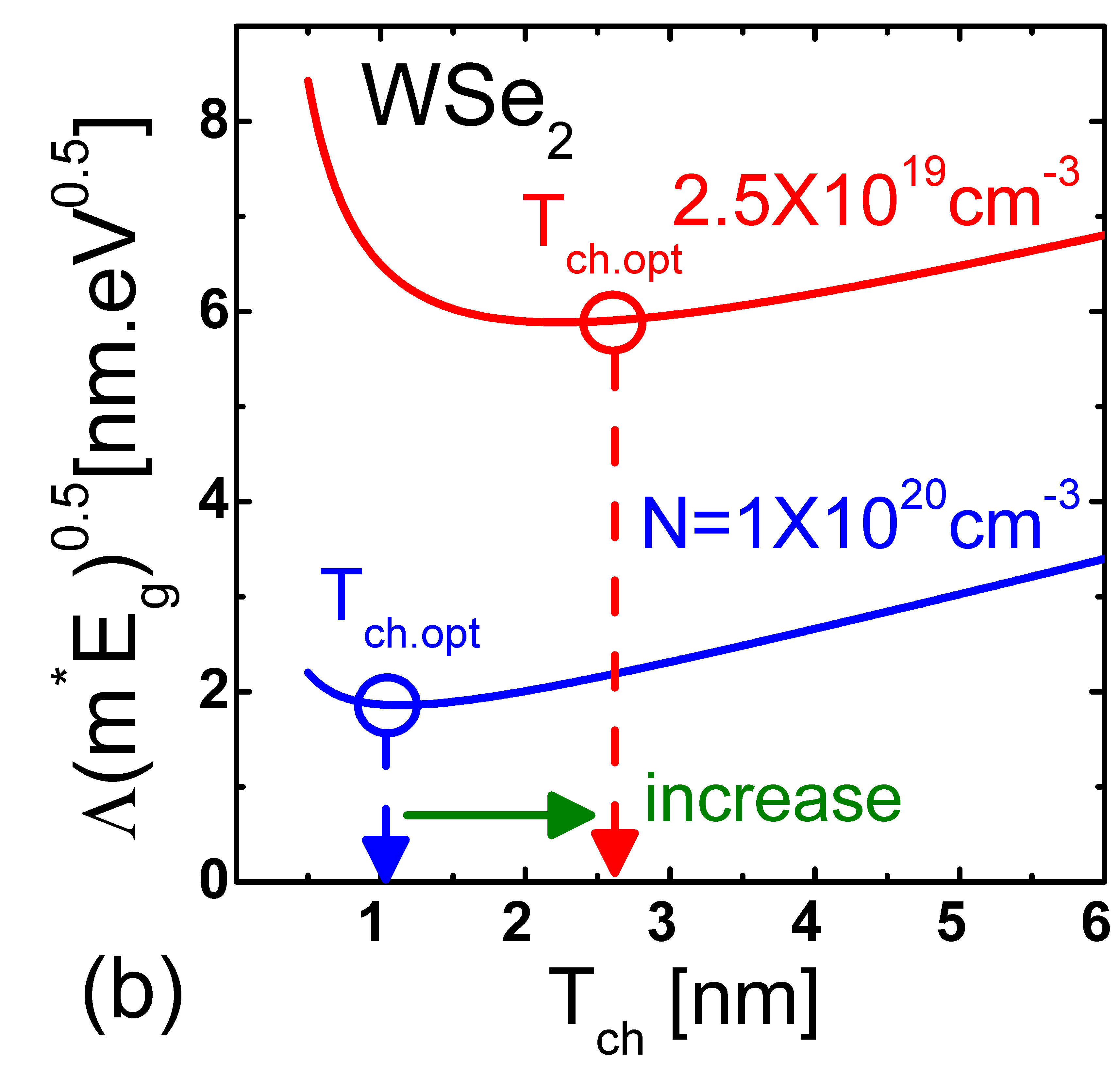}
\caption{(a) $\Lambda\sqrt{m^* E_{g}}$ for WSe$_{2}$, BP, and InAs TFETs at a source doping density of $10^{20} cm^{-3}$. The solid lines are calculated from eq. (\ref{eq:LmEg_for_opt_tch}) with parameters $c_{0} \sim c_{5}$ as listed in Appendix III. The stars are extracted from atomistic quantum transport simulations. (b) $\Lambda\sqrt{m^* E_{g}}$ of WSe$_{2}$ TFETs for two different source doping densities calculated from eq. (\ref{eq:LmEg_for_opt_tch}). $T_{ch.opt}$ increases as the doping density is reduced. }
\label{LmEg}
\end{figure}

The optimum $T_{ch}$ ($T_{ch.opt}$) for the ON-state minimizes $\Lambda\sqrt{m^* E_{g}}$ and is expected to maximize $I_{ON}$. The dependence of $\Lambda$, $m^*$, and $E_{g}$ on $T_{ch}$ has been discussed in section II and III. All of them can be expressed as a function of $T_{ch}$. As a result, $\Lambda\sqrt{m^* E_{g}}$ can also be expressed as a function of $T_{ch}$ and is given by
\begin{equation}
\begin{aligned} 
& \Lambda\sqrt{m^* E_{g}} \\
& = (\lambda + W_{D}) \sqrt{m^* E_{g}} \\
& = \left( c_0 T_{ch} + c_1 + c_2 \dfrac{1}{T_{ch}} + c_3 \dfrac{1}{T_{ch}^{2}} \right)c_4  \sqrt{1+c5\dfrac{1}{T_{ch}}}  \label{eq:LmEg_for_opt_tch}        
\end{aligned}
\end{equation}

parameters $c_{0} \sim c_{5}$ are described in Appendix III in detail.
Finding $T_{ch.opt}$ that minimizes $\Lambda\sqrt{m^* E_{g}}$ can be accomplished analytically or numerically. An exact analytic $T_{ch.opt}$ solved by $\frac{d\Lambda \sqrt{m^*E_{g}}}{dT_{ch}}=0$ is complicated to interpret. Therefore, we will focus in the following on the numerical results by calculating $\Lambda \sqrt{m^*E_{g}}$ at different $T_{ch}$ and find $T_{ch.opt}$ as the minimum of those plots. Fig. \ref{LmEg} (a) shows $\Lambda \sqrt{m^*E_{g}}$  for WSe$_{2}$, BP, and InAs TFETs calculated form eq. (\ref{eq:LmEg_for_opt_tch}), corresponding well with the results from atomistic quantum simulations. 

It is apparent that WSe$_{2}$ as a class I material exhibits a smaller $T_{ch.opt}$ as mentioned before, since $\frac{\alpha}{E_{g.bulk}}<<$ 0.5 nm which is a single atomic layer's thickness. For a class II material like BP or InAs, $\frac{\alpha}{E_{g.bulk}}$ is larger than a mono-layer's thickness which implies a larger $T_{ch.opt}$. Moreover, $T_{ch.opt}$ increases when the source doping density (N) decreases as apparent from fig. \ref{LmEg}(b), since $W_{D}$ inversely proportionally depends on N as stated above.

\begin{table} [!h]
\begin{center}
\rule{0pt}{8pt}
\begin{tabular}{ |c|ccc| } 
 \hline
         &            &              &             \\ [-0.8em]
         & WSe$_{2}$  & BP           & InAs  \\ 
         &            &              &         \\  [-1em]
 \hline
          &            &             &         \\  [-0.8em]
$\alpha/E_{g.bulk}$ [nm]   & 0.01  &  1.8 & 4.3 \\
          &            &      &         \\  [-1em]
\hline
          &            &      &         \\  [-0.8em]
Class     & I          &  II  & II       \\ 
 \hline
\end{tabular}
\end{center}
\caption{Classification of WSe$_{2}$, BP, and InAs. }
\label{table:Eg_m}
\end{table}

\section{An upper limit of $T_{ch}$ defined by the OFF-state ($T_{ch.OFF}$)}

\begin{figure}[!t]
\centering
\includegraphics[width=1.6in]{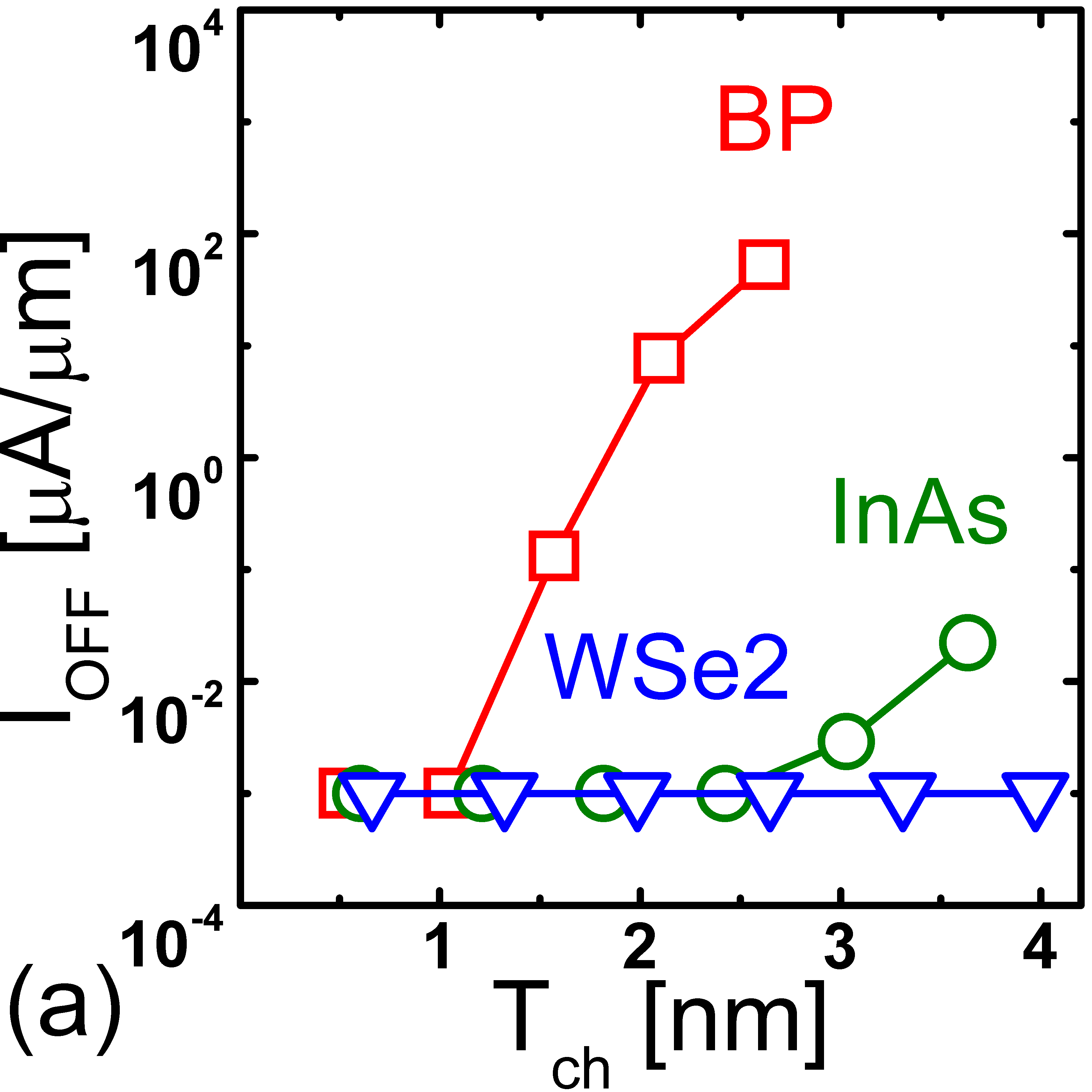}
\includegraphics[width=1.6in]{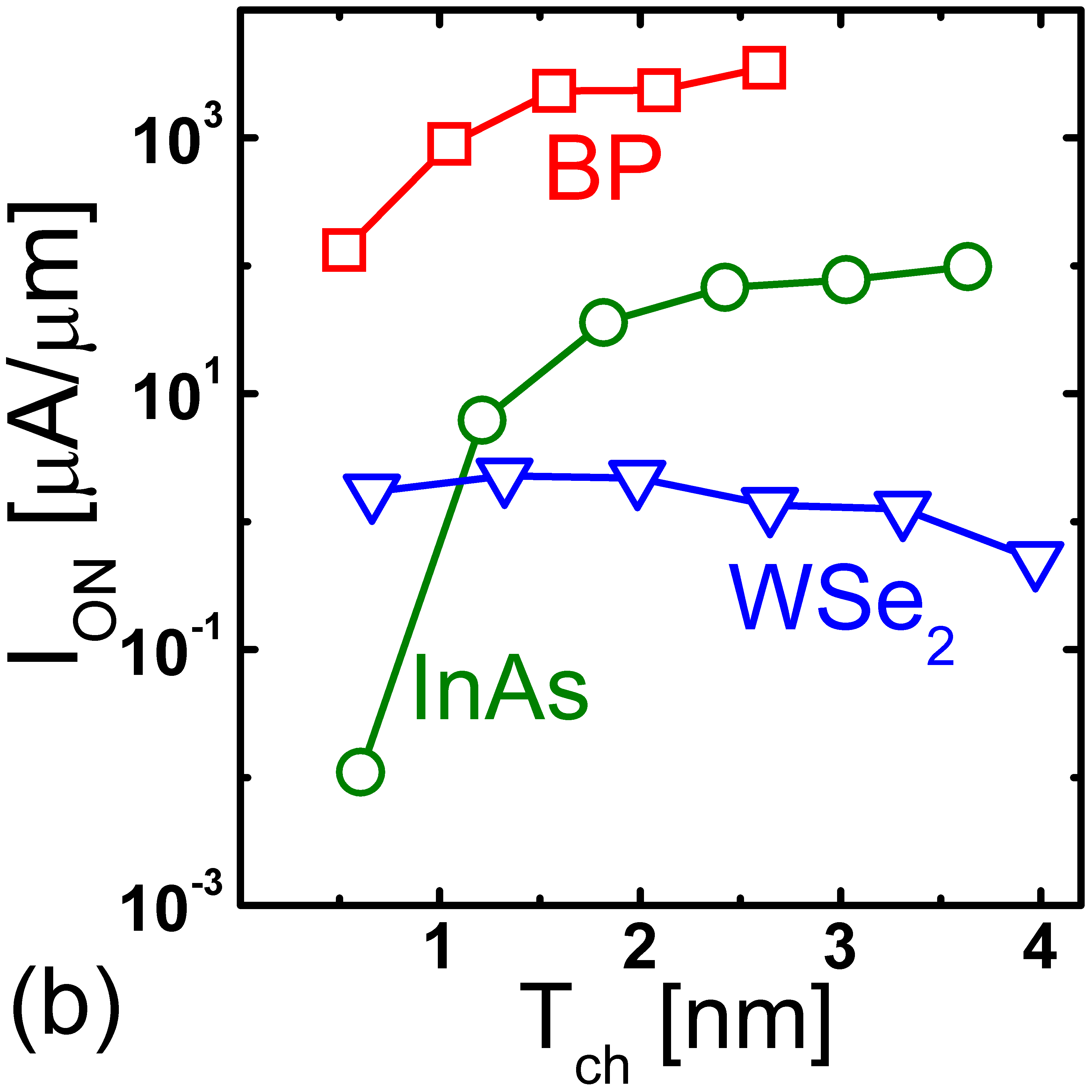}\\
\includegraphics[width=1.6in]{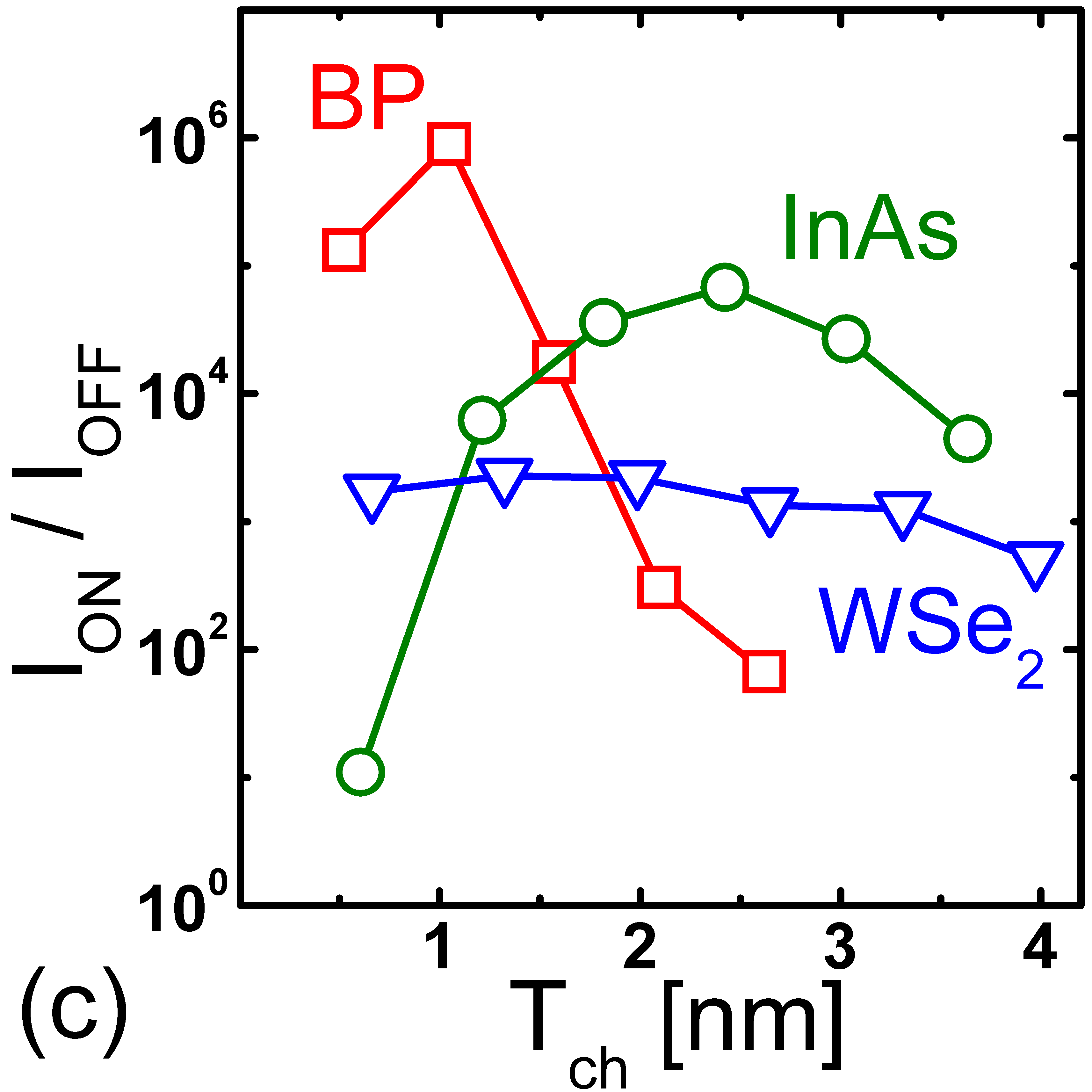}
\includegraphics[width=1.6in]{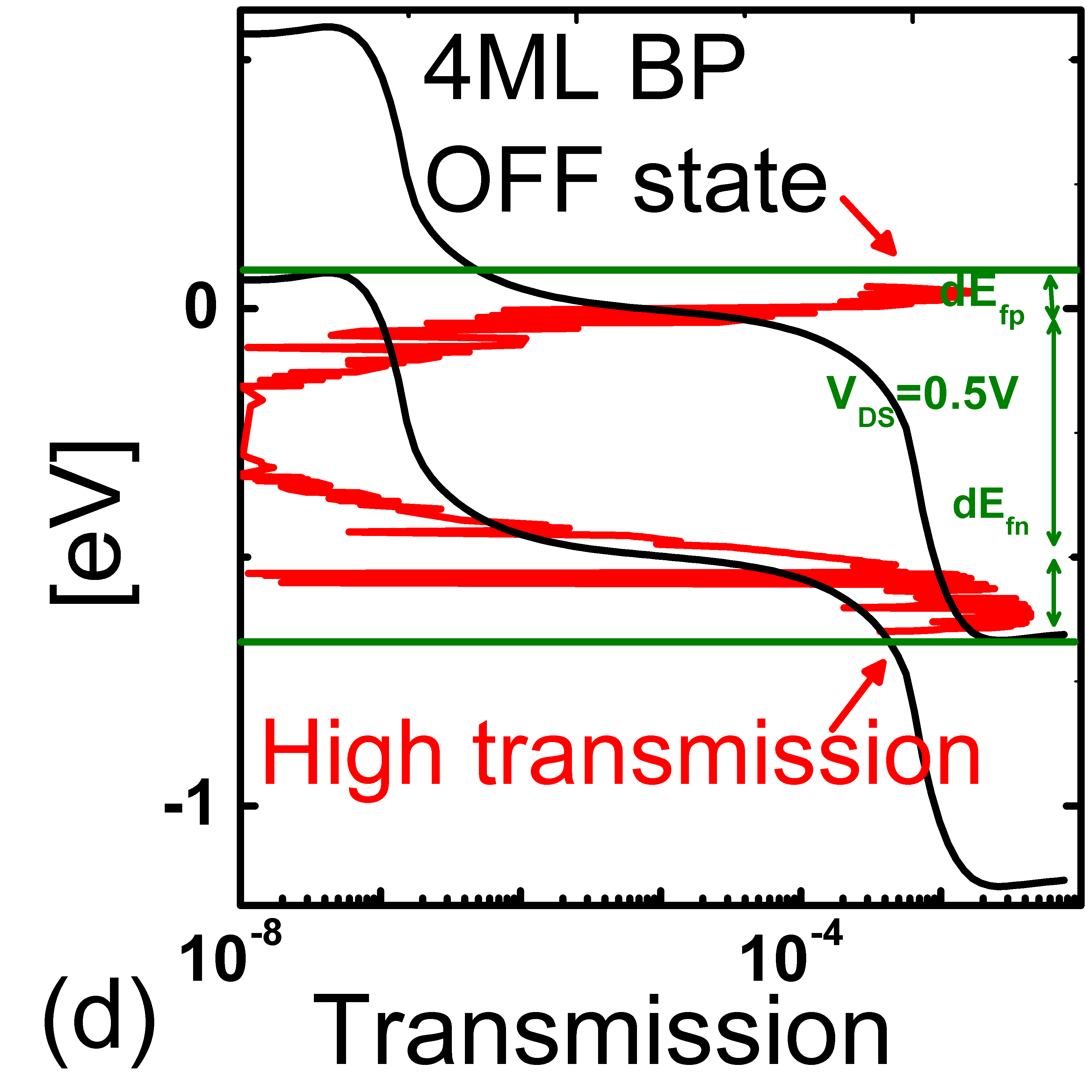}

\caption{(a), (b), and (c) Dependence of $I_{OFF}$, $I_{ON}$, and the ON/OFF current ratio on $T_{ch}$. (d) OFF-state transmission for a BP 4ML TFET. A significant source to drain leakage in thick $T_{ch}$ BP (and InAs) TFETs is due to the small $E_{g}$. }
\label{ON_OFF_ratio_CD-TFETs}
\end{figure}

A thicker channel with a smaller bandgap may increase $I_{OFF}$ and deteriorate the ON/OFF current ratio. An upper limit on $T_{ch}$ can be deduced by considering the maximum permissible OFF current ($I_{OFF}$). Fig. \ref{ON_OFF_ratio_CD-TFETs}(a), (b), and (c) show the $I_{OFF}$, $I_{ON}$, and ON/OFF current ratio extracted from the $I_{D}-V_{G}$ curves as calculated from atomistic quantum transport simulations. The simulated structure is shown in Appendix II and assumes the supply voltage ($V_{DD}$) is 0.5V. The $I_{D}-V_{G}$ is shifted by adjusting the OFF-state ($I_{D}=I_{OFF}$) to $V_{G}=0$ V. $I_{ON}$ is extracted at $V_{G}=V_{D}=0.5$ V after shifting the $I_{D}-V_{G}$ curves. Following the convention of the ITRS roadmap \cite{itrs2015}, $I_{OFF}$ is chosen to be $10^{-3} \mu A/\mu m$ or the minimum possible current above this value. $I_{OFF}$ for BP and InAs is substantial when $T_{ch}$ is above 1 nm and 3 nm respectively. This is because $E_{g}$ in this case is too small to block the tunneling current in the OFF-state as shown in the Fig. \ref{ON_OFF_ratio_CD-TFETs}(d).

The ON/OFF current ratio suffers from significant degradation if the device cannot be turned off effectively as shown in Fig. \ref{ON_OFF_ratio_CD-TFETs}(c). This would occur if $E_{g} < qV_{DS}+dE_{fp}+dE_{fn}$. $dE_{fp}$ and $dE_{fn}$ are the difference between the Fermi level and the band edge in the degenerately doped source and drain region respectively. There exist a $T_{ch.OFF}$ below which $E_g$ is large enough to suppress the OFF current. Given that, $E_{g}=E_{g.bulk}+\dfrac{\alpha}{T_{ch}}$, $T_{ch.OFF}$ can be expressed as
\begin{align} \label{eq:max_Tch} 
  T_{ch.OFF} & < | \dfrac{\alpha}{qV_{DS}-E_{g.bulk}+dE_{fp}+dE_{fn}}|
\end{align}  
$dE_{fn}$ and $dE_{fp}$ reduces as the doping density decreases, which results in a larger $T_{ch.OFF}$.

\begin{figure}[!t]
\centering
\includegraphics[width=1.6in]{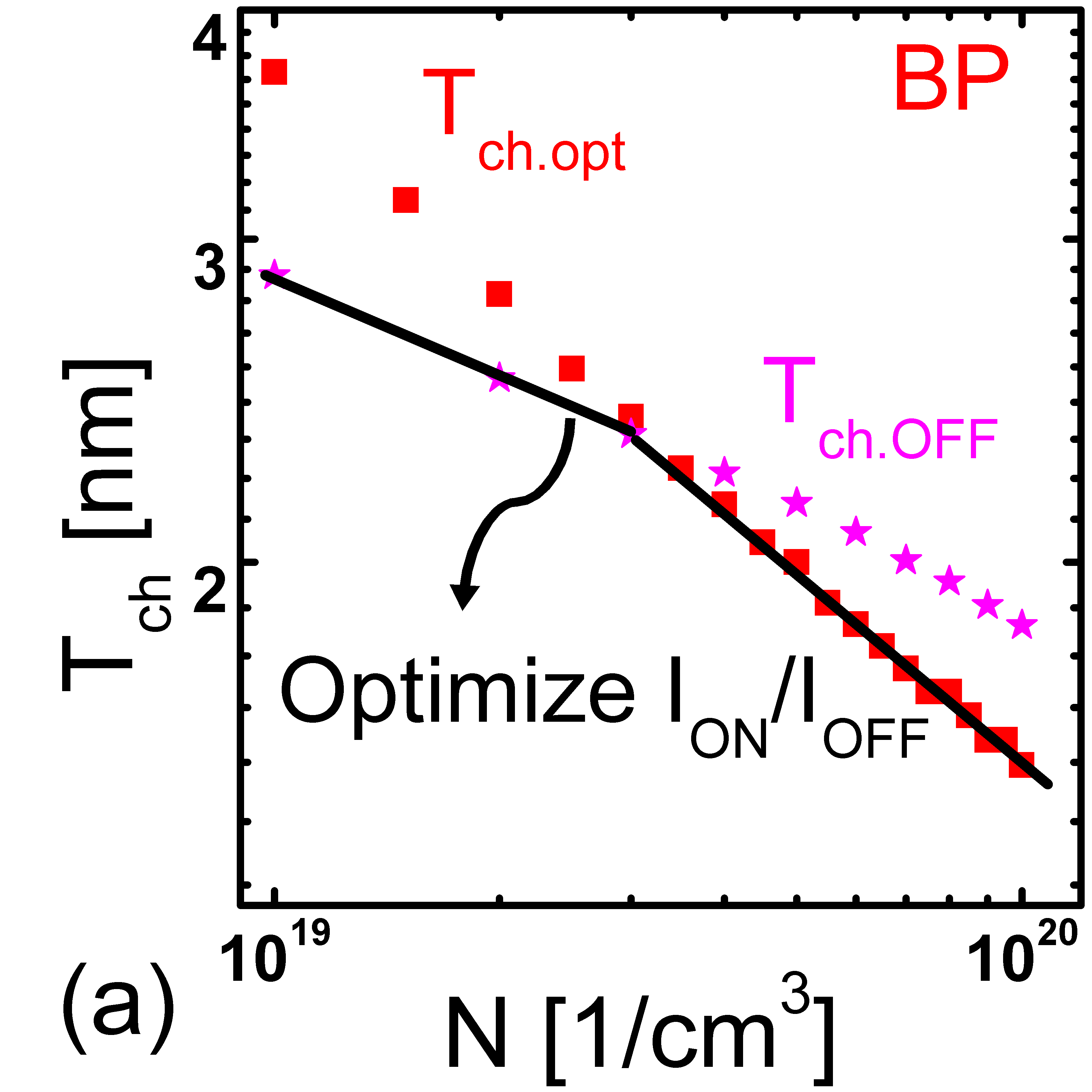}
\includegraphics[width=1.6in]{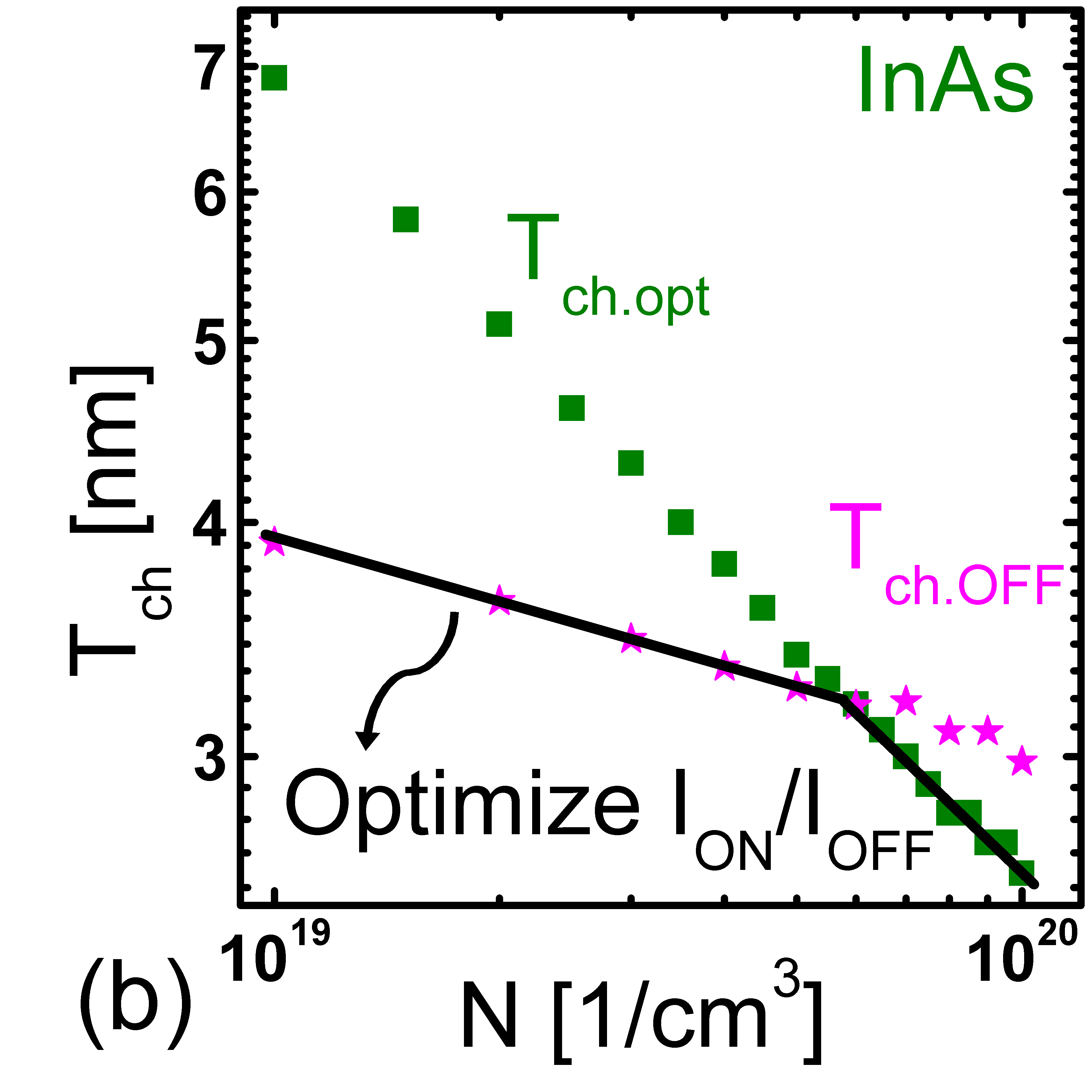}
\caption{(a) and (b) $T_{ch.opt}$ and $T_{ch.OFF}$ for different source doping densities. The channel thickness that optimizes the ON/OFF current ratio is $T_{ch.opt}$ or $T_{ch.OFF}$ whichever is smaller; $min(T_{ch.opt},T_{ch.OFF})$. Note that $T_{ch.OFF}$ is obtained from atomistic quantum transport simulations for $V_{DS}$=0.5V.}
\label{optimum_tch}
\end{figure}

To optimize a TFET's ON/OFF current ratio, $T_{ch.opt}$ or $T_{ch.OFF}$ whichever is smaller should be used. Fig. \ref{optimum_tch} (a) and (b) show how $T_{ch.opt}$ and $T_{ch.OFF}$ for BP and InAs change as a function of the source doping density. Both $T_{ch.opt}$ and $T_{ch.OFF}$ increase as the source doping density is reduced.

\section{Summary}
Optimizing the channel thickness of a 2D TFET can significantly improve its performance. The choice of the channel thickness affects both the material properties and the electrostatics. There exists a channel thickness that minimizes the tunneling distance. However, the ON-state channel thickness ($T_{ch.opt}$) should optimize the product of the band gap, reduced effective mass and square of the tunneling distance. Moreover, a maximum permissible channel thickness ($T_{ch.OFF}$) is needed to reach acceptable OFF-currents. In this work, compact models were introduced to describe these two important channel thicknesses. A 2D TFET exhibits the highest ON/OFF current ratio when the channel thickness is chosen to be the smaller of $T_{ch.opt}$ and $T_{ch.OFF}$.

\vspace{5mm} 
\textbf{Appendix I. How descriptive is $I_{ON} \propto e^{- \Lambda \sqrt{m^{*}E_{g}}}$ ?}
\vspace{5mm} 

\begin{figure}[!t]
\centering
\includegraphics[width=2.0in]{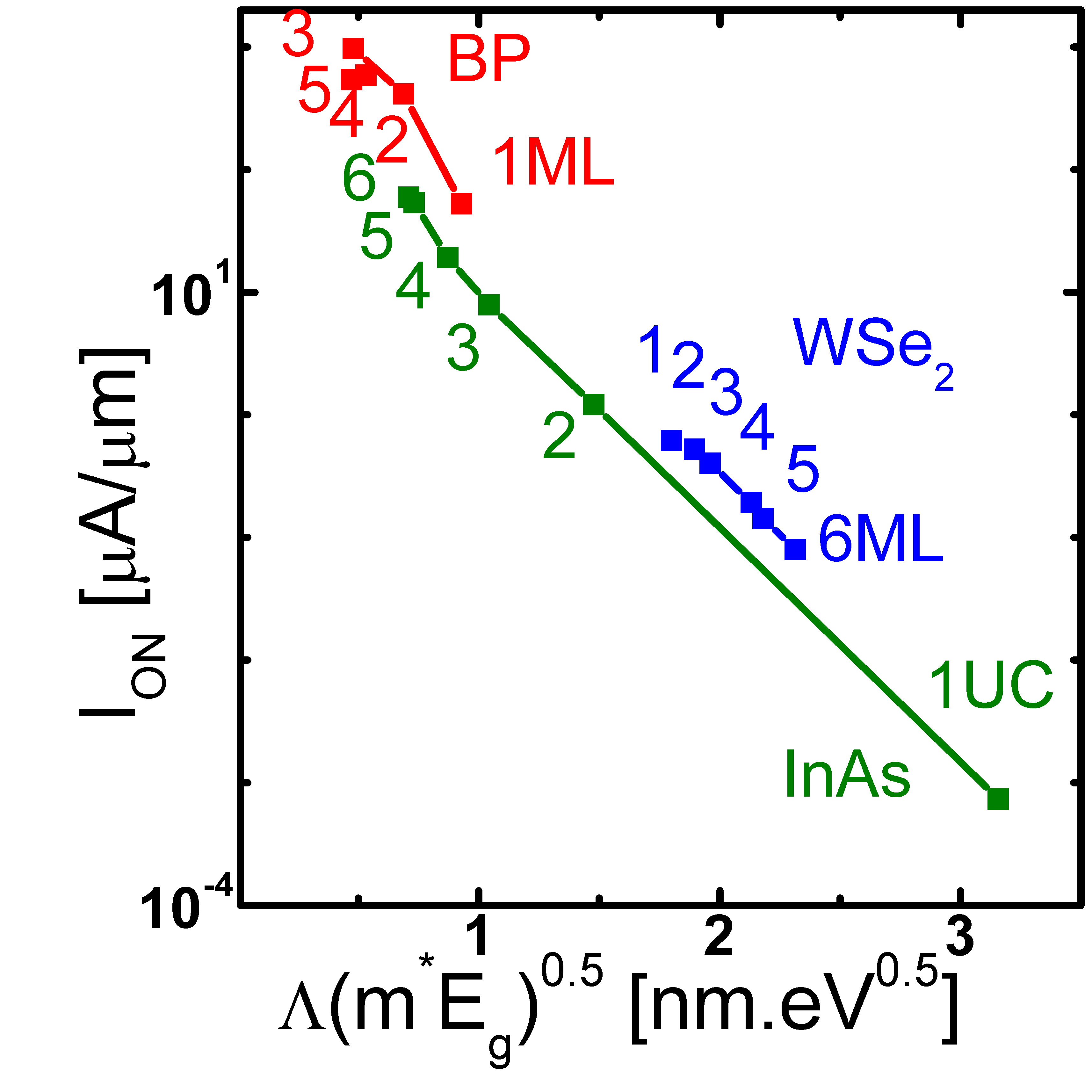}
\caption{ $I_{ON}$ and its corresponding $\Lambda\sqrt{m^{*}E_{g}}$ are extracted from atomistic simulations. $I_{ON}$ is proportional to $e^{-\Lambda \sqrt{m^{*}E_{g}}}$. ML and UC are the abbreviations for mono-layer and unit cell respectively.}
\label{benchmark_Ion}
\end{figure}

Knowing how descriptive is $I_{ON} \propto e^{- \Lambda \sqrt{m^{*}E_{g}}}$ compared to sophisticated quantum transport simulations is critical before optimizing $T_{ch}$ by minimizing $\Lambda \sqrt{m^{*}E_{g}}$. Fig. \ref{benchmark_Ion} shows  $I_{ON}$ and its corresponding $\Lambda \sqrt{m^{*}E_{g}}$ from full band self consistent atomistic simulations. 

This compact equation quantitatively represents the trend of sophisticated atomistic simulations, although it assumes a simple potential distribution and a simple $m^*$ \cite{Hesam_FN_tunneling}, Different materials that have the same $\Lambda \sqrt{m^{*}E_{g}}$ are expected to provide $I_{ON}$ within the same order of magnitude. However, we find minor deviations between different materials because this compact equation approximates complex band structures by a single band reduced effective mass and ignores contributions from higher sub-bands. To demonstrate the descriptiveness of this compact equation, $I_{on}$ and $\Lambda$ in Fig. \ref{benchmark_Ion} are extracted when the tunneling window is opened to 0.35 eV to reduce contributions from higher sub-bands.

\vspace{5mm} 
\textbf{Appendix II. Simulation Details}
\vspace{5mm} 

\begin{figure}[!b]
\centering
\includegraphics[width=3.0in]{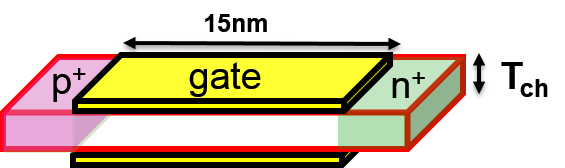}
\caption{Structure of the chemically doped TFETs.}
\label{CD_TFET_structure}
\end{figure}

A schematic structure of the simulated double gated chemically doped TFET is shown in Fig. \ref{CD_TFET_structure}. The channel is 15 nm with the supply voltage ($V_{DD}=0.5V$) following the ITRS 15nm technology node \cite{ITRS_roadmap}. The oxide is assumed to be H$_{F}$O$_{2}$ with an EOT of 0.5 nm. The source is heavily doped at density $\sim 10^{20}$ 1/$cm^3$.


The numerical simulations are performed self-consistently by coupling quantum trasmissting boundary method (QTBM) and 3D-Poisson equation \cite{NEGF_method}. 3D finite-difference method is used to calculate the carrier density ($\rho$). The anisotropic dielectric constant is taken into account in the Poisson equation as shown in eq. (\ref{eq:Poisson}) \cite{Wang2015,Kumar2012}.

\begin{align} \label{eq:Poisson}
\dfrac{d}{dx} \left( \varepsilon_{in} \dfrac{dV}{dx} \right) + \dfrac{d}{dy} \left( \varepsilon_{in} \dfrac{dV}{dy} \right) + \dfrac{d}{dz} \left( \varepsilon_{out} \dfrac{dV}{dz} \right) = -\rho 
\end{align}

where $\varepsilon_{in}$ and $\varepsilon_{out}$ are the in-plane and out-of-plane dielectric constants. The QTBM method is equivalent to the nonequalibrium Green's function approach without scattering but is more computationally efficient \cite{Hesam_TFETs_2D_TMD}. The open boundary Schr$\ddot{o}$dinder equation is solved in the following form:

\begin{align}
(EI-H-\Sigma)\psi=S
\end{align}

where $E$, $I$, $H$, and $\Sigma$ are energy, identity matrix, device Hamiltonian, and the total self-energy due to the open boundaries condition. $\psi$ and $S$ are the wave function in the device and the strengh of the carrier injection from contacts respectively. The Hamiltonian is constructed with the second nearest neighbor tight binding method. The simulation is performed by the Nanoelectronics Modeling tool: NEMO5. \cite{NEMO5_1,NEMO5_2} More simulation details can be found in \cite{Ameen2016} and \cite{Hesam_sensitivity}.


\vspace{5mm} 
\textbf{Appendix III Analytical expression for $\Lambda \sqrt{m^{*}E_{g}}$ }
\vspace{5mm} 

$\Lambda \sqrt{m^{*}E_{g}}$ can be expressed as a function of $T_{ch}$ as 
\begin{align}   
\Lambda\sqrt{m^* E_{g}} & = (\lambda + W_{D}) \sqrt{m^* E_{g}}   \\         
& = \left( c_0 T_{ch} + c_1 + c_2 \dfrac{1}{T_{ch}} + c_3 \dfrac{1}{T_{ch}^{2}} \right) c_4 \sqrt{1+c5\dfrac{1}{T_{ch}}}  \label{eq:LmEg_for_opt_tch_apdix2}        
\end{align} 
where $c_{0}$ through $\sim c_{5}$ are obtained after expressing $\lambda$, $W_{D}$, $m^*$, and $E_{g}$ as a function of $T_{ch}$. The dependence of $m^*$ and $E_{g}$ on $T_{ch}$ has been discussed in Section II. 

$W_{D}$ for a 2D PN-junction is given by
\begin{align}              
W_{D} & = \dfrac{\pi \varepsilon \Delta V}{ln(4)qNT_{ch}}  \label{eq:WD1_apdix}\\
& \sim \dfrac{\pi \varepsilon E_{g}}{ln(4)qNT_{ch}}  \label{eq:WD2_apdix}\\
& \sim \dfrac{\pi \left(\varepsilon_{1} T_{ch} +\varepsilon_{2}\right) \left( E_{g.bulk}+\dfrac{\alpha}{T_{ch}}\right)}{ln(4)qNT_{ch}} \label{eq:WD3_apdix}           
\end{align}
which can be expressed as a function of $T_{ch}$ explicitly. $\Delta V$ is of the order of $E_{g}$ which can be expressed as $E_{g}=E_{g.bulk}+\dfrac{\alpha}{T_{ch}}$. $\varepsilon$ increases with $T_{ch}$ due to changes in the electrostatic environment and can be expressed as $\varepsilon \sim \varepsilon_{1} T_{ch} +\varepsilon_{2}$. $\varepsilon_{1}$ and $\varepsilon_{2}$ can be obtained by fitting $W_{D}$ shown in Fig. \ref{tunnel_distance}(a), (b), and (c). 

\begin{table} [!htbp]
\begin{center}
\rule{0pt}{8pt}
\begin{tabular}{ |c|cc| } 
 \hline
 & &  \\ [-1em]
 channel    & $\varepsilon_{1}$    &  $\varepsilon_{2}$   \\ 
            & [$\varepsilon_{0}$]  & [$\varepsilon_{0}/nm$] \\ 
 \hline
  & & \\ [-1em]
 WSe$_{2}$  & 0.43     &  0.99    \\ 
  & &  \\ [-1em]
 BP         & 0.62    &  0.91     \\ 
  & & \\ [-1em]
 InAs       & 0.83    &  1.28    \\ 
 \hline
\end{tabular}
\end{center}
\caption{Parameters $\varepsilon_{1}$ and $\varepsilon_{2}$ for WSe$_{2}$, BP, and InAs. $\varepsilon_{0}=8.854 \times  10^{-12}$ C/(V.m).}
\label{table:eps}
\end{table}


On the other hand, the chemically doped TFET's scaling length ($\lambda$) is given by 
\begin{align}    
\lambda & =  \dfrac{\varepsilon}{\pi \varepsilon_{ox}}\left[ \gamma_1 T_{ch} + \gamma_2 T_{ox} \right]\label{eq:lambda1_apendix}\\
&= L_{1} T_{ch} + L_{2}.
\end{align} 

which is a modified version of the electrically doped TFETs' $\lambda$ \cite{Hesam_DETFET, Hesam_ED_TFET}. $L_{1}$ and $L_{2}$ can be obtained by fitting $\lambda$ shown in Fig. \ref{tunnel_distance} (a), (b), and (c). 

    
\begin{table} [!htbp]
\begin{center}
\rule{0pt}{8pt}
\begin{tabular}{ |c|cc| } 
 \hline
 & &  \\ [-1em]
 channel    & $L_{1}$    &  $L_{2}$   \\ 
            & [cons.]  & [nm] \\ 
 \hline
  & & \\ [-1em]
 WSe$_{2}$  & 0.7     &  0.12    \\ 
  & &  \\ [-1em]
 BP         & 1.13   &  -0.46     \\ 
  & & \\ [-1em]
 InAs       & 0.94    &  0.35    \\ 
 \hline
\end{tabular}
\end{center}
\caption{Parameters $L_{1}$ and $L_{2}$ for WSe$_{2}$, BP, and InAs.}
\label{table:L1_L2}
\end{table}

After substituting $W_{D}$, $\lambda$, $E_{g}$, and $m^*$, $\Lambda \sqrt{m^*E_{g}}$ can be rearranged as
\begin{align*}   
     \Lambda\sqrt{m^* E_{g}}  = (\lambda + W_{D}) \sqrt{m^* E_{g}}  
\end{align*}
\begin{align*}  
= \left( L_{1}T_{ch}+L_{2} + \dfrac{\pi \left(\varepsilon_{1} T_{ch} +\varepsilon_{2}\right) \left( E_{g.bulk}+\dfrac{\alpha}{T_{ch}}\right)}{ln(4)qNT_{ch}}\right)
\end{align*}
\begin{align*}
\hspace{130pt} \sqrt{m^* \left( E_{g.bulk}+\dfrac{\alpha}{T_{ch}}\right)} 
\end{align*}
\begin{align}   
& = \left( c_0 T_{ch} + c_1 + c_2 \dfrac{1}{T_{ch}} + c_3  \dfrac{1}{T_{ch}^{2}} \right) c_4  \sqrt{1+c5\dfrac{1}{T_{ch}}} \label{eq:Lambda_appendix_part3}        
\end{align}
where $c_{1}$ to $c_{5}$ are      
\begin{align}  
          c_{0} & = L_{1} \label{eq:c0}\\
          c_{1} & = L_{2} + \beta_{1}E_{g.bulk} \label{eq:c1} \\
          c_{2} & = \beta_{2}E_{g.bulk}+\beta_{1}\alpha  \label{eq:c2}\\
          c_{3} & = \beta_{2}\alpha \label{eq:c3}\\
          c_{4} & = \sqrt{m^* E_{g.bulk}} \label{eq:c4}\\ 
          c_{5} & = \dfrac{\alpha}{E_{g.bulk}} \label{eq:c5} \\      
      \beta_{1} & = \dfrac{\pi \varepsilon_{1}}{ln(4)qN}  \label{eq:beta1}\\
      \beta_{2} & = \dfrac{\pi \varepsilon_{2}}{ln(4)qN}   \label{eq:beta2}  
      \end{align}   
$c_{1}$ to $c_{5}$ for the three channel materials are listed in the Table \ref{table:c0_c5}. 
\begin{table} [!htbp]
\begin{center}
\begin{tabular}{ |c|cccccccc| } 
 \hline
  & & & & & & & & \\ [-1em]
 channel    & $c_{0}$ & $c_{1}$ & $c_{2}$ &  $c_{3}$ & $c_{4}$ & $c_{5}$ & $\beta1$ & $\beta2$ \\ 
   & & & & & & & & \\ [-1em]
\hline
    & & & & & & & & \\ [-1em]
 WSe$_{2}$  & 0.7     &  1.78   & 0.85    &  0.01    & 0.554    & 0.01   & 1.24    &  0.54 \\ 
    & & & & & & & & \\ [-1em]

    & & & & & & & & \\ [-1em]
 BP         & 1.14    &  -0.24   & 0.71    &  0.57    & 0.12     & 1.79   &  0.78    &  1.14 \\ 
    & & & & & & & & \\ [-1em]
 
    & & & & & & & & \\ [-1em]
 InAs       & 0.94    &  0.91   & 2.77    &  1.56    & 0.09     & 4.28   &  1.60    &  1.04 \\ 

 \hline
\end{tabular}
\end{center}
\caption{Parameters $c_1$ to $c_5$ for WSe$_{2}$, BP, and InAs.}
\label{table:c0_c5}
\end{table}



\bibliographystyle{ieeetr}

\bibliography{tch_optimization_citation}
\end{document}